\documentclass[showpacs,prl,twocolumn,floatfix]{revtex4}
\usepackage{graphicx,amssymb,amsmath}

\begin{document}

\title{Emergence of hierarchy in cost driven growth of spatial networks}

\author{Remi Louf}
\email{remi.louf@cea.fr}
\affiliation{Institut de Physique Th\'{e}orique, CEA, CNRS-URA 2306, F-91191, 
Gif-sur-Yvette, France}

\author{Pablo Jensen}
\email{pablo.jensen@ens-lyon.fr}
\affiliation{IXXI, Rh\^one Alpes Institute of complex systems, 69364 Lyon, France}
\affiliation{Laboratoire de Physique, UMR CNRS 5672, ENS de Lyon, 69364 Lyon, France}

\author{Marc Barthelemy}
\email{marc.barthelemy@cea.fr}
\affiliation{Institut de Physique Th\'{e}orique, CEA, CNRS-URA 2306, F-91191, 
Gif-sur-Yvette, France}
\affiliation{CAMS (CNRS/EHESS) 190-198, avenue de France, 75244 Paris Cedex 13, France}

\begin{abstract} 

  One of the most important features of spatial networks such as transportation networks, power grids, Internet, neural networks, is the existence of a cost associated with the length of links. Such a cost has a profound influence on the global structure of these networks which usually display a hierarchical spatial organization. The link between local constraints and large-scale structure is however not elucidated and we introduce here a generic model for the growth of spatial networks based on the general concept of cost benefit analysis. This model depends essentially on one single scale and produces a family of networks which range from the star-graph to the minimum spanning tree and which are characterised by a continuously varying exponent. We show that spatial hierarchy emerges naturally, with structures composed of various hubs controlling geographically separated service areas, and appears as a large-scale consequence of local cost-benefit considerations. Our model thus provides the first building blocks for a better understanding of the evolution of spatial networks and their properties. We also find that, surprisingly, the average detour is minimal in the intermediate regime, as a result of a large diversity in link lengths. Finally, we estimate the important parameters for various world railway networks and find that -- remarkably -- they all fall in this intermediate regime, suggesting that spatial hierarchy is a crucial feature for these systems and probably possesses an important evolutionary advantage. 

\end{abstract}

\maketitle

\section{Introduction}

Our societies rely on various networks for the distribution of energy, information and for transportation of individuals. These networks shape the spatial organization of our societies and their understanding is a key step towards the understanding of the characteristics and the evolution of our cities~\cite{Batty:2005}. Despite their apparent diversity, these networks are all particular examples of a broader class of networks --spatial networks-- which are characterised by the embedding of their nodes in space. As a consequence, there is usually a cost associated with a link, leading to particular structures which are now fairly well understood~\cite{Barthelemy:2011}, thanks to the recent availability of large sets of data. Nevertheless, the mechanisms underlying the formation and temporal evolution of spatial networks have not been much studied. Different kinds of models aiming at explaining the static characteristics of spatial networks have been suggested previously in quantitative geography, transportation economics, and physics (for a review, see ~\cite{Levinson:2009}). Concerning the time evolution of spatial networks, a few models only exist to describe in particular the growth of road and rail networks~\cite{Levinson:2006, Gastner:2006, Barthelemy:2008, Courtat:2011, Black:1971}, but a general framework is yet to be discovered.

The earliest attempts can be traced back to the economic geography community in the 60s and 70s (A fairly comprehensive review of these studies can be found in~\cite{Haggett:1969}). However, due to the lack of available data and computational power, most of the proposed models were based on intuitive, heuristic rules and have not been studied thoroughly.

A more recent trend is that of the optimization models. The common point between all these models is that they try to reproduce the topological features of existing networks, by considering the network as the realisation of the optimum of given quantity (see section IV.E in~\cite{Barthelemy:2011} for an overview). For instance, the hub-and-spoke models~\cite{OKelly:1998} reproduce correctly with an optimization procedure the observed hierarchical organization of city pair relations. However, the vast majority of the existing spatial networks do not seem to result from a global optimization, but rather from the progressive addition of nodes and segments resulting from a local optimization. By modeling (spatial) networks as resulting from a global optimization, one overlooks the usually limited time horizon of planners and the self-organization underlying their formation.

Self-organization of transportation networks has already been studied in transportation engineering~\cite{Levinson:2006, Xie:2009}. Using an agent-based model including various economical ingredients, the authors of~\cite{Levinson:2006} modeled the emergence of the networks properties as a degeneration process. Starting from an initial grid, traffics are computed at each time step and each edge computes its costs and benefits accordingly, using any excess to improve their speed. After several iterations, a hierarchy of roads emerges. Our approach is very different: we start from nodes and we do not specify any initial network. Also, and most importantly, we deliberately do not represent all the causal mechanisms at work in the system. Indeed, the aim of our model is to understand the basic ingredients for emergence of patterns that can be observed in various systems and we thus focus on a single, very general economical mechanism and its consequence on the large-scale properties of the networks.

Concerning spatial networks, as it is the case for many spatial structure, there is a strong path dependency. In other words, the properties of a network at a certain time can be explained by the particular historical path leading to it. It thus seems reasonable to model spatial networks in an iterative way. Some iterative models, following ideas for understanding power laws in the Internet \cite{Fabrikant:2002} and describing the growth of transportation networks~\cite{Gastner:2006} can be found in the literature. In these models, the graphs are constructed via an iterative greedy optimization of geometrical quantities. However, we believe that the topological and geometrical properties of networks are consequences of the underlying processes at stake. At best, geometrical and topological quantities can be a proxy for other --more fundamental-- properties: for instance, it will be clear in what follows that the length of an edge can be taken as a proxy for the cost associated with the existence of that edge. Finding those underlying processes is a key step towards a general framework within which the properties of networks can be understood and, hopefully, predicted.

In this respect, cost-benefit analysis (CBA) provides a systematic method to evaluate the economical soundness of a project. It allows one to appreciate whether the costs of a decision will outweigh its benefits and therefore evaluate quantitatively its feasibility and/or suitability. Cost-benefit analysis has only been officially used to assess transport investments since 1960~\cite{Coburn:1960}. However, the concept comes accross as so intuitive in our profit-driven economies that it seems reasonable to wonder whether CBA is at the core of the emergent features of our societies such as distribution and transportation systems. If the temporal evolution of spatial networks is rarely studied, arguments mentioning the costs and benefits related to such networks are almost absent from the physics litterature (\cite{Popovic:2011} is a notable exception, although they do not consider the time evolution of the network.). However, we find it intuitively appealing that in an iterative model, the formation of a new link should --at least locally-- correspond to a cost-benefit analysis. We therefore propose here a simple cost-benefit analysis framework for the formation and evolution of spatial networks. Our main goal within this approach is to understand the basic processes behind the self-organization of spatial networks that lead to the emergence of their large scale properties.

\section{Theoretical formulation}

We consider here the simple case where all the nodes are distributed uniformly in the plane (see Methods for detailed description of the algorithm). For a rail network, the nodes would correspond to cities and the network grows by adding edges between cities iteratively; the edges are added sequentially to the graph --as a result of a cost-benefit analysis-- until all the nodes are connected~\cite{Black:1971}. For the sake of simplicity, we limit ourselves to the growth of trees which allows to focus on the emergence of large-scale structures due to the cost-benefit ingredient alone.  Furthermore, we consider that all the actors involved in the building process are perfectly rational and therefore that the most profitable edge is built at each step. More precisely, at each time step we build the link connecting a new node $i$ to a node $j$ which already belongs to the network, such that the following quantity is maximum
\begin{equation}
\label{eq:general_framework}
R_{ij} = B_{ij} - C_{ij}
\end{equation}
The quantity $B_{ij}$ is the \textit{expected benefit} associated with the construction of the edge between node $i$ and node $j$ and $C_{ij}$ is the \textit{expected cost} associated with such a construction. Eq. (\ref{eq:general_framework}) defines the general framework of our model and we now discuss specific forms of $R_{ij}$. In the case of transportation networks, the cost will essentially correspond to some maintenance cost and will typically be proportional to the euclidean distance $d_{ij}$ between $i$ and $j$. We thus write
\begin{equation}
\label{eq:cost}
C_{ij} = \kappa d_{ij}
\end{equation}
where $\kappa$ represents the cost of a line per unit of length per unit of time. Benefits are more difficult to assess. For rail networks, a simple yet reasonable assumption is to write the benefits in terms of distance and expected traffic $T_{ij}$ between cities $i$ and $j$ 
\begin{equation} 
\label{eq:benefits} 
B_{ij} = \eta T_{ij} d_{ij}
\end{equation}
where $\eta$ represents the benefits per passenger per unit of length. We have to estimate the expected traffic between two cities and for this we will follow the common and simple assumption used in the transportation litterature, of having the so-called gravity law ~\cite{Stewart:1948,Erlander:1990}
\begin{equation} 
\label{eq:gravity} 
T_{ij} = k\frac{M_i \: M_j}{d_{ij}^a} 
\end{equation}
where $M_{i(j)}$ is the population of city $i(j)$, and $k$ is the rate associated with the process. We will choose here a value of the exponent $a>1$ ($a<1$ would correspond to an unrealistic situation where the benefits associated with passenger traffic would increase with the distance). This parameter $a$ determines the range at which a given city attracts traffic, regardless of the density of cities. The accuracy and relevance of this gravity law is still controversial and improvements have been recently proposed~\cite{Simini:2012,Gargiulo:2012}. But it has the advantage of being simple and to capture the essence of the traffic phenomenon: the decrease of the traffic with distance and the increase with population. Within these assumptions, the cost-benefit budget $R'_{ij}=R_{ij}/\eta$ now reads
\begin{equation}
\label{eq:R0}
R'_{ij} = k\frac{M_i M_j}{d_{ij}^{\;a-1}} - \: \beta d_{ij}
\end{equation}
where $\beta = \frac{\kappa}{\eta}$ represents the relative importance of the cost with regards to the benefits. We will assume that populations are power-law distributed with exponent $\mu$ (which for cities is approximatively $\mu\approx 1.1$, see Methods) and the model thus depends essentially on the two parameters $a$, and $\beta$ (for a detailed description of parameter used in this paper, see Methods section). In the following we will be working with fixed values of $\mu$ and $a$. The exact values we choose are however not important (as long as their are chosen in a given range, see previous section and the Materials and Methods section) as the obtained graphs would have the same qualitative properties.

\section{Results}

\subsection{Typical scale} The average population is $\overline{M}$ and the typical inter-city distance is given by $\ell_1\sim 1/\sqrt{\rho}$ where $\rho=N/L^2$ denotes the city density ($L$ is the typical size of the whole system). The two terms of Eq.~\ref{eq:R0} are thus of the same order for $\beta=\beta^*$ defined as

\begin{equation}
\label{eq:beta*}
\beta^* = k \overline{M}^2 \rho^{a/2}
\end{equation}

In the theoretical discussion that follows, we will take $k=1$ for simplicity (but it should not be forgotten in empirical discussions). Another way of interpreting $\beta^*$ which makes it more practical to estimate from empirical data (see section Discussion), is to say that it is of the order of the average traffic per unit time
\begin{equation}
\label{eq:beta*_traffic}
\beta^* = < T >
\end{equation}

From Eq.~\ref{eq:beta*} we can guess the existence of two different regimes depending on the value of $\beta$:
\begin{itemize}
\item $\beta \ll \beta^*$ the cost term is negligible compared to the benefits term. Each connected city has its own influence zone depending on its population and the new cities will tend to connect to the most influent city. In the case where $a\approx 1$, every city connects to the most populated cities and we obtain a star graph constituted of one single hub connected to all other cities.
\item $\beta \gg \beta^*$ the benefits term is negligible compared to the cost term. All new cities will connect sequentially to their closest neighbour. Our algorithm is then equivalent to an implementation of Prim's algorithm~\cite{Prim:1957}, and the resulting graph is a minimum spanning tree (MST).
\end{itemize}

The intermediate regime $\beta\simeq\beta^*$ however needs to be elucidated.  In particular, we have to study if there is a transition or a crossover between the two extreme network structures, and if we have a crossover what is the network structure in the intermediate regime. In the following we answer these questions by simulating the growth of these spatial networks.

\subsection{Crossover between the star graph and the MST}

\begin{figure}[!h]
\centering
\includegraphics[width=0.5\textwidth]{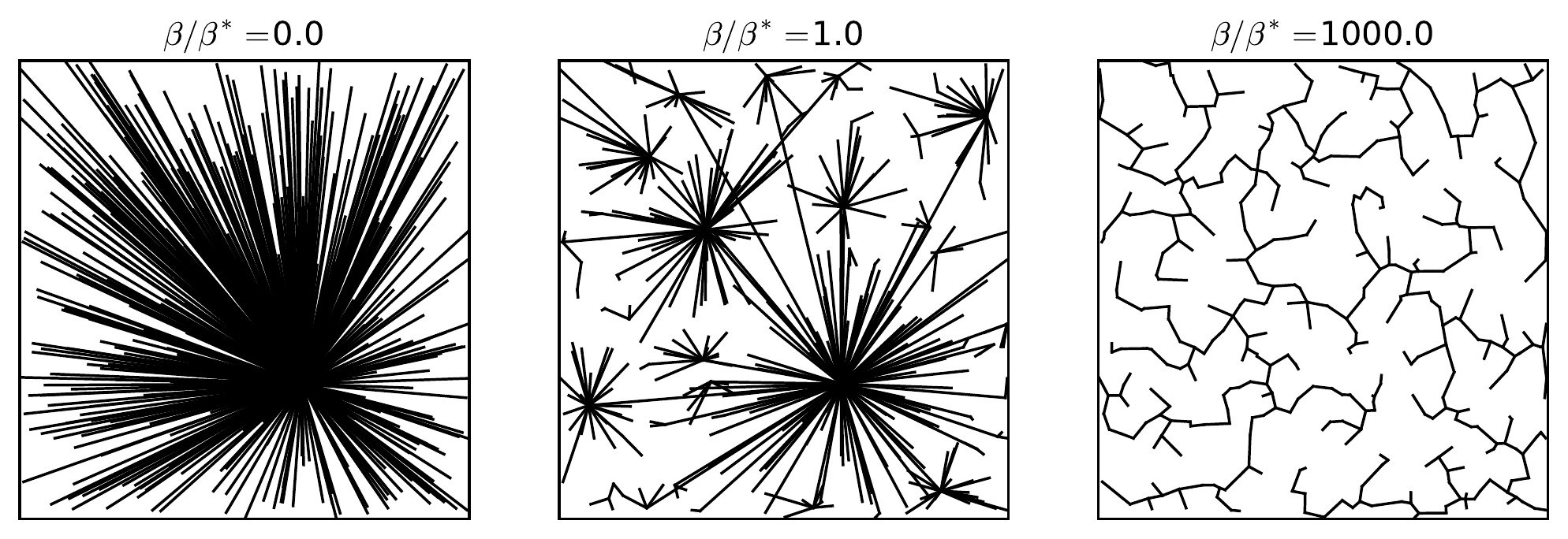}
\caption{Graphs obtained with our algorithm for the same set of cities (nodes) for three different values of $\beta^*$ ($a=1.1$, $\mu=1.1$, $400$ cities). On the left panel, we have a star graph where the most populated node is the hub and on the right panel, we recover the minimum spanning tree.
\label{fig:plot_graphs}}
\end{figure}

Fig.~\ref{fig:plot_graphs} shows three graphs obtained for the same set of cities for three different values of $\beta/\beta^*$ ($a=1.1$, $\mu=1.1$) confirming our discussion about the two extreme regimes in the previous section. A visual inspection seems to show that for $\beta \sim \beta^*$ a different type of graph appears, which suggests the existence of a crossover between the star-graph and the MST. This graph is reminiscent of the hub-and-spoke structure that has been used to describe the interactions between city pairs~\cite{OKelly:1998,OKelly:1996}. However, in contrast with the rest of the literature about hub-and-spoke models, we show that this structure is not necessarily the result of a global optimization: indeed, it emerges here as the result of the auto-organization of the system.

\begin{figure}[!h]
\centering
\includegraphics[width=0.50\textwidth]{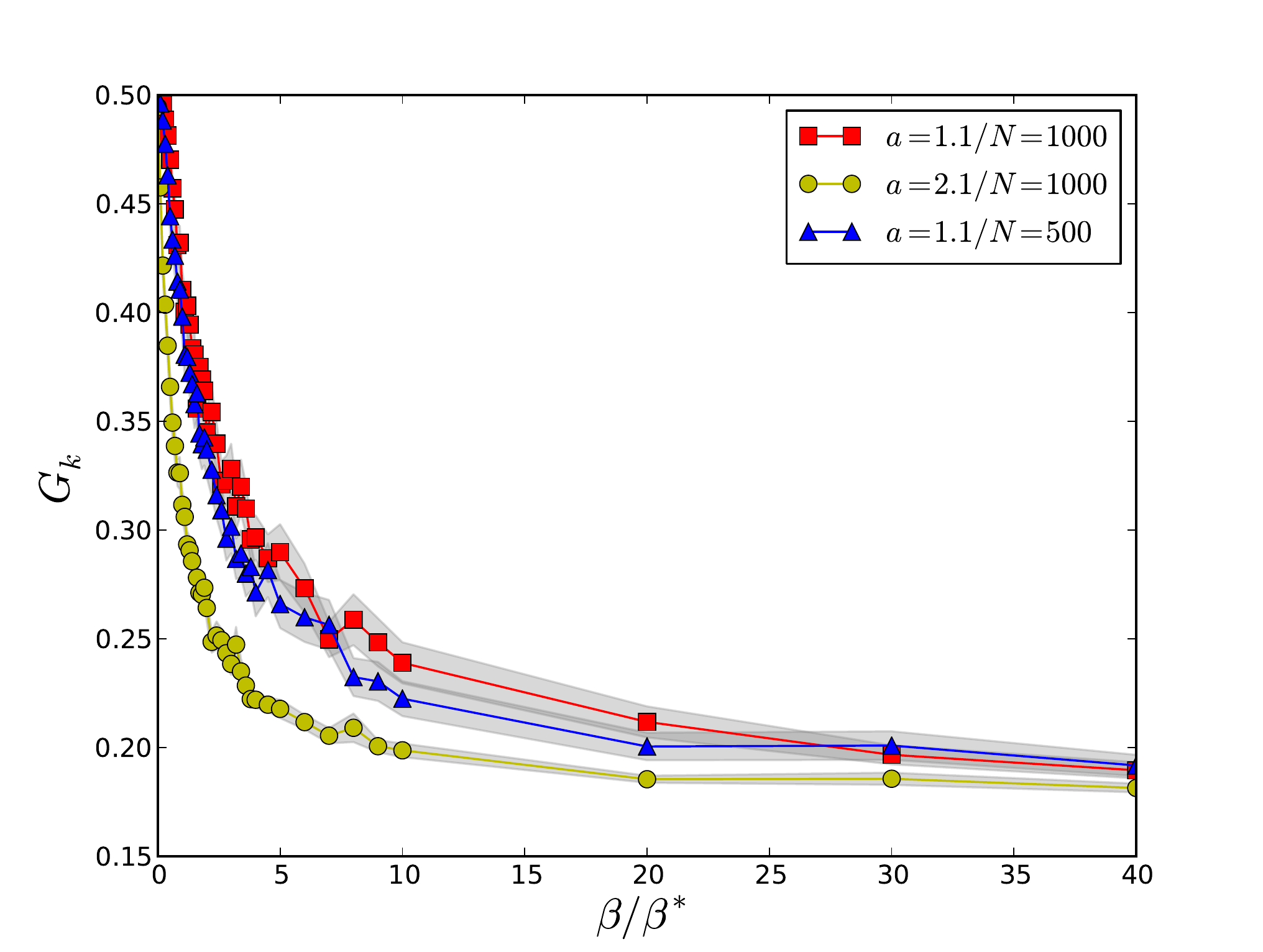}
\caption{Evolution of the Gini coefficient with $\beta/\beta^*$ for different values of $\beta^*$. The shaded area represents the standard deviation of the Gini coefficient.\label{fig:gini}}
\end{figure}

The MST is characterised by a peaked degree distribution while the star graph's degree distribution is bimodal, and we therefore choose to monitor the crossover with the Gini coefficient for the degrees defined as in~\cite{Gini:1987}
\begin{equation}
\label{eq:gini}
G_k = \frac{1}{2 N^2 \bar{k}} \sum_{i,j=1}^{N} | k_i - k_j |
\end{equation}
where $\bar{k}$ is the average degree of the network. The Gini coefficient is in $[0,1]$ and if all the degrees are equal, it is easy to see that $G=0$. On the other hand, if all nodes but one are of degree 1 (as in the star-graph), a simple calculation shows that $G=1/2$. Fig.~\ref{fig:gini} displays the evolution of the Gini coefficient versus $\beta/\beta^*$ (for different values of $\beta^*$ obtained by changing the value of $a$, $\mu$ and $N$). This plot shows a smooth variation of the Gini coefficient pointing to a crossover between a star graph and the MST,  as one could expect from the plots on Fig.~\ref{fig:plot_graphs} (also, we note that for given values of $a$, $\mu$ all the plots collapse on the same curve, regardless of the number $N$ of nodes. However for different values of $a$ or $\mu$ we obtain different curves).

\begin{figure}[!h]
\centering
\includegraphics[width=0.50\textwidth]{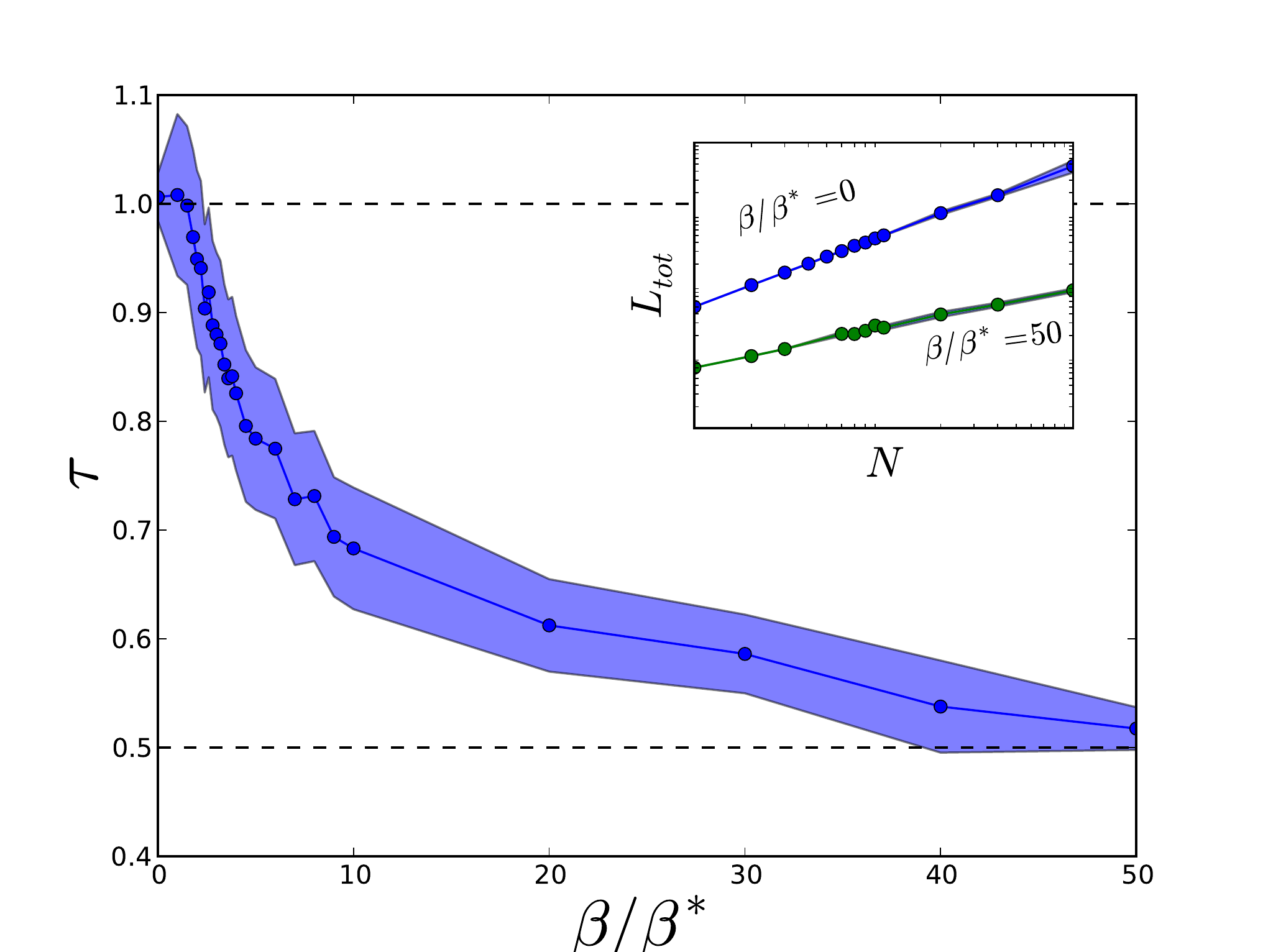} 
\caption{Exponent $\tau$ versus $\beta$. For $\beta\ll \beta^*$ we recover the star-graph exponent $\tau=1$ and for the other extreme $\beta\gg\beta^*$ we recover the MST exponent $\tau=1/2$. In the intermediate range, we observe a continuously varying exponent suggesting a non-trivial structure. The shaded area represents the standard deviation of $\tau$.\label{fig:Ltot_vs_beta}. Inset: In order to illustrate how we determined the value of $\tau$, we represent $L_{tot}$ versus $N$ for two different values of $\beta$. The power law fit of these curves gives $\tau$.}
\end{figure}

Another important difference between the star-graph and the MST lies in how the total length of the graph scales with its number of nodes. Indeed, in the case of the star-graph, all the nodes are connected to the same node and the typical edge length is $L$, the typical size of the system the nodes are enclosed in. We thus obtain
\begin{equation}
\label{eq:Ltot_star}
L_{tot} \sim L\; N
\end{equation}
On the other hand, for the MST each node is connected roughly to its nearest neighbour at distance typically given by $\ell_1\sim L/\sqrt{N}$, leading to
\begin{equation}
\label{eq:Ltot_MST}
L_{tot} \sim L\; \sqrt{N}
\end{equation}
More generally, we expect a scaling of the form $L_{tot}\sim N^\tau$ and on Fig.~\ref{fig:Ltot_vs_beta} we show the variation of the exponent $\tau$ versus $\beta$. For $\beta=0$ we have $\tau=1.0$ and we recover the behavior $L_{tot} \propto N$ typical of a star graph. In the limit $\beta \gg \beta^*$ we also recover the scaling $L_{tot} \propto \sqrt{N}$, typical of a MST. For intermediate values, we observe an exponent which varies continuously in the range $[0.5,1.0]$. This rather surprising behavior is rooted in the heterogeneity of degrees and in the following, we will show that we can understand this behaviour as resulting from the hierarchical structure of the graphs in the intermediate regime. 

It is interesting to note that a scaling with an exponent $1/2<\tau < 1$ has been observed~\cite{Samaniego:2008,Barthelemy:2011} for the total number $\ell_T$ of miles driven by the population (of size $P$) of city scales as $\ell_T \propto P^{\beta}$ with $\beta=0.66$. Understanding the origin of those intermediate numbers might thus also give us insights into important features of traffic in urban areas and the structure of cities.

It thus seems that from the point of view of interesting quantities such as the Gini coefficient or the exponent $\tau$, there is no sign of a critical value for $\beta$ and that we are in presence of a crossover and not a transition.

\subsection{Spatial Hierarchy and scaling}

The graph corresponding to the intermediate regime $\beta \approx \beta^*$ depicted on Fig.~\ref{fig:plot_graphs} exhibits a particular structure corresponding to a hierarchical organization, observed in many complex networks~\cite{Sales-Pardo:2007}. Inspired from the observation of networks in the regime $\beta/\beta^* \sim 1$, we define a particular type of hierarchy --that we call \emph{spatial hierarchy}-- as follows. A network will be said to be spatially hierarchical if:

\begin{enumerate}
\item  We have a hierarchical network of hubs that connect to nodes less and less far away as one goes down the hierarchy;
\item Hubs belonging to the same hierarchy level have their own influence zone clearly separated from the others'. In addition, the influence zones of a given level are included in the influence zones of the previous level.
\end{enumerate} 

\begin{figure}[!h]
\centering
\includegraphics[width=0.50\textwidth]{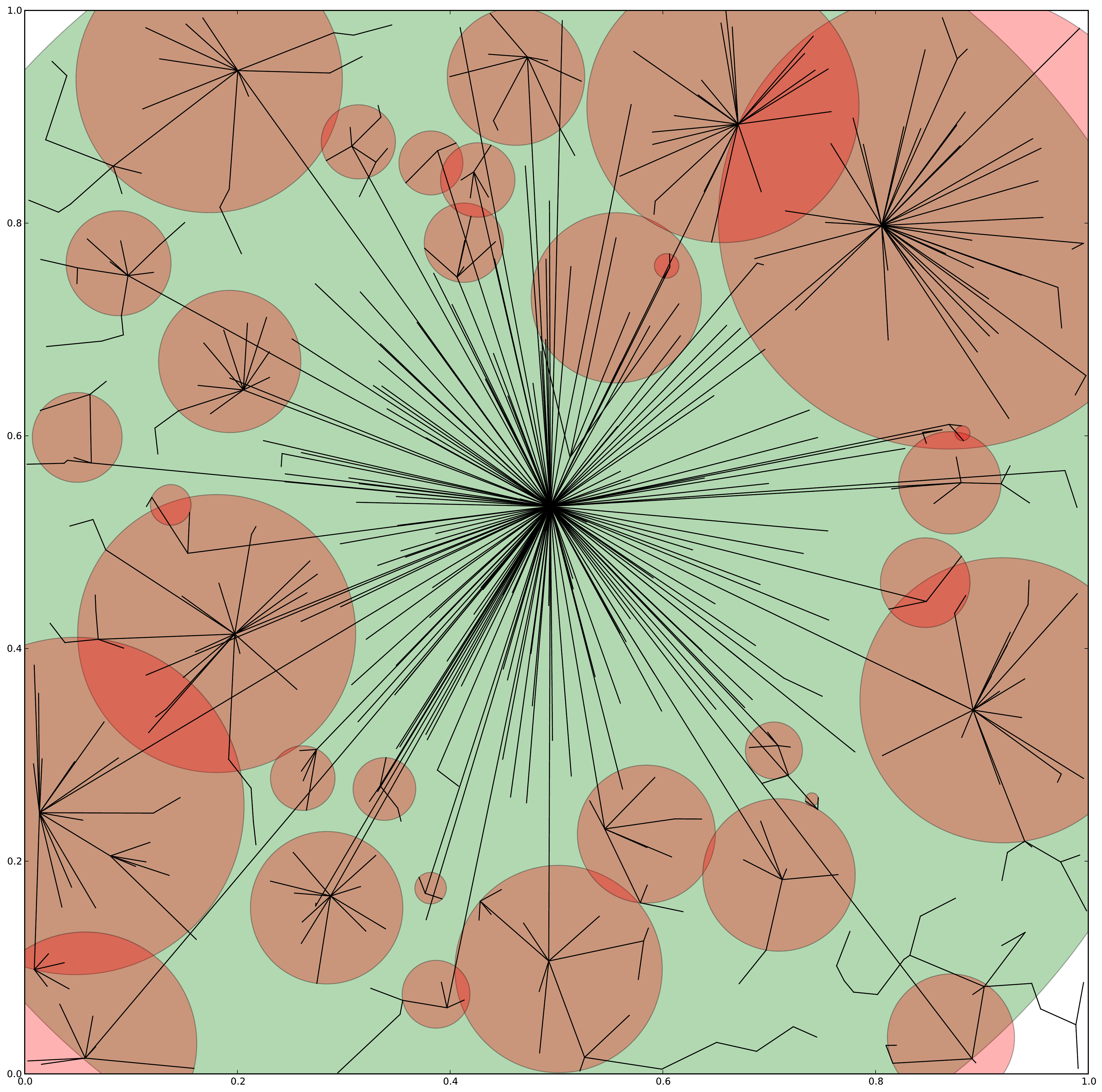} 
\caption{Example of a graph where we represent the influence zones for the first two hierarchical levels.\label{fig:separation_example}}
\end{figure}

The relevance of this new concept of hierarchy in the present context can be qualitatively assessed on Fig.~\ref{fig:separation_example} where we represent the influence zones by colored circles, the colors corresponding to different hierarchical levels. In order to go beyond this simple, qualitative description of the structure, we provide in the following a quantitative proof that networks in the regime $\beta / \beta^*$ exhibit spatial hierarchy.

\subsubsection{Distance between hierarchical levels}
 
\begin{figure}[!h]
\centering
\includegraphics[width=0.5\textwidth]{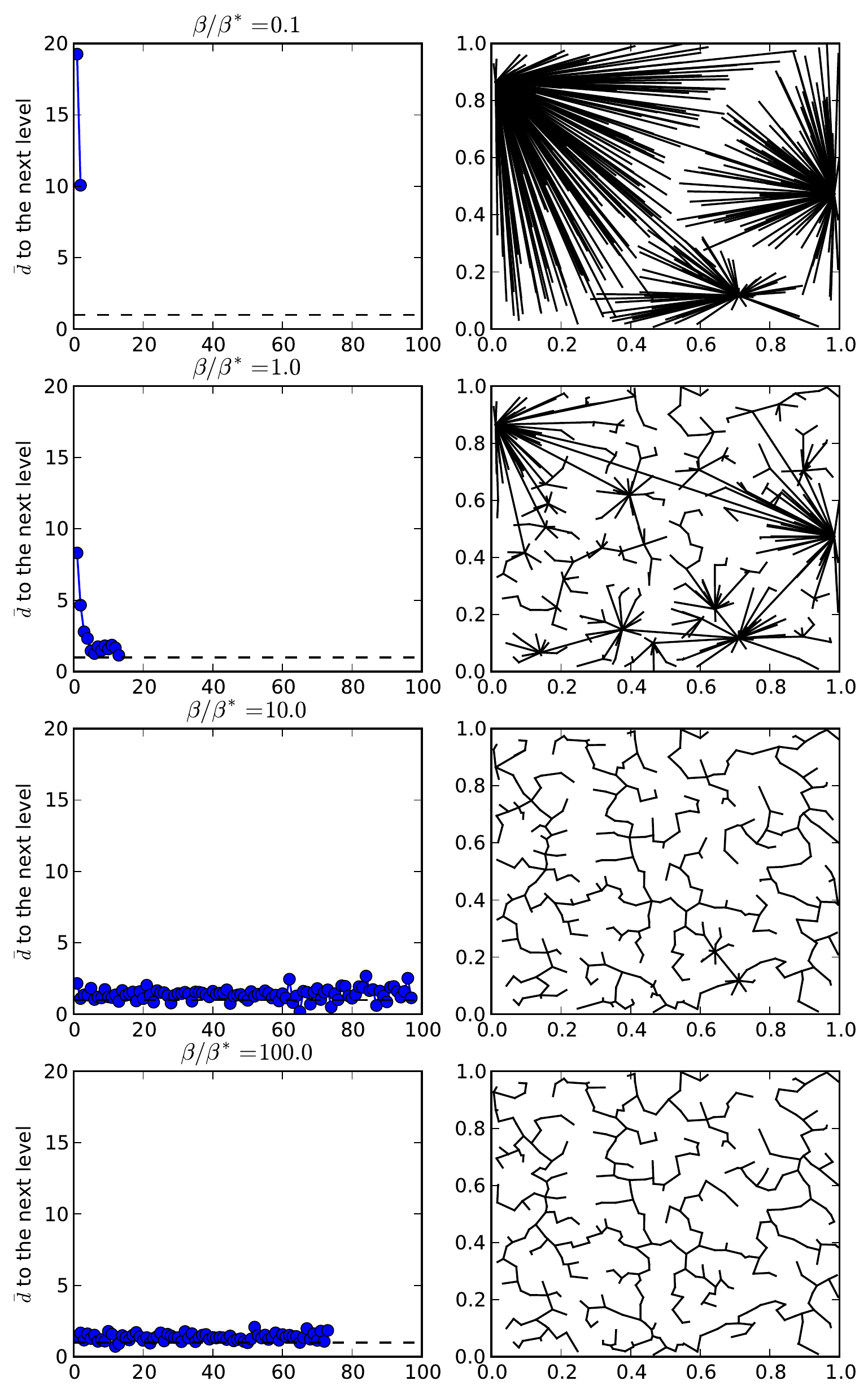}
\caption{Left column: Average distance between the successive hierarchy levels for different values of $\beta/\beta^*$, next to the corresponding graphs (on the right column). The most populated node is taken as the root node. \label{fig:distance_hierarchy}}
\end{figure}

We propose here a quantitative characterisation of the part (1) in the definition of spatial hierarchy. The first step is to identify the root of the network which allows us to naturally characterize a hierarchical level by its topological distance to the root. We choose the most populated node as the root (which will be the largest hub for $\beta\ll\beta^*$) and we can now measure various quantities as a function of the level in the hierarchy. In Fig.~\ref{fig:distance_hierarchy}, we plot the average euclidean distance $\overline{d}$ between the different hierarchical levels as a function of the topological distance from the root node (for the sake of clarity, we also draw next to these plots the corresponding graphs). For reasonably small  values of $\beta/\beta^*$ (i.e. when the graph is not far from being a star-graph), the average distance between levels decreases as we go further away from the root node. This confirms the idea that the graphs for $\beta/\beta^*\simeq 1$ exhibit a spatial hierarchy where nodes from different levels are getting closer and closer to each other as we go down the hierachy. Eventually, as $\beta/\beta^*$ becomes larger than 1, the distance between consecutive levels just fluctuates around $\ell_1\sim 1/\sqrt{\rho}$ the average distance between nearest neighbours for a Poisson process, which indicates the absence of hierarchy in the network.

\subsubsection{Geographical separation of hubs zones} 

\begin{figure}[!h]
\centering
\includegraphics[width=0.30\textwidth]{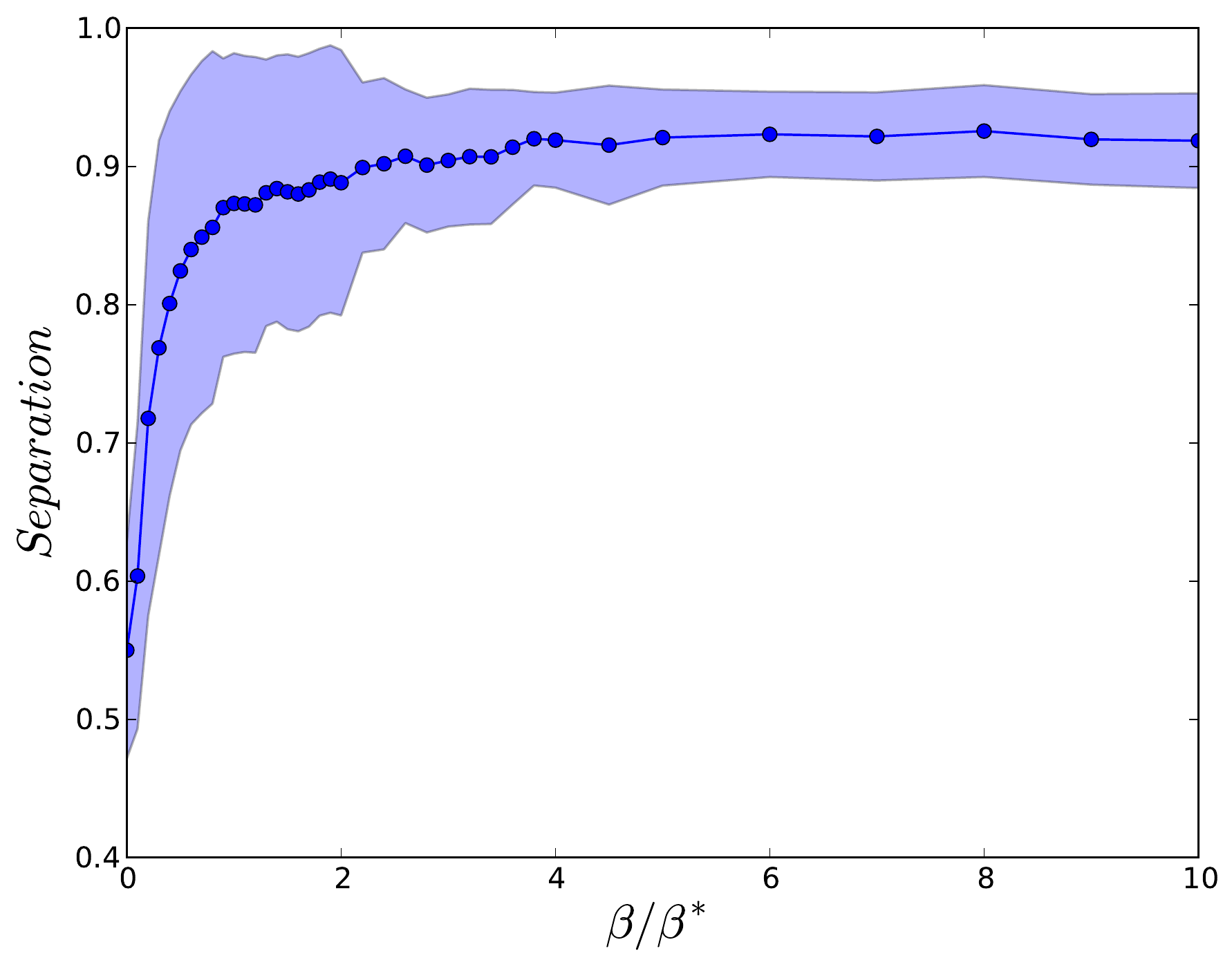}
\caption{Separation index averaged over all the graph's level versus $\beta/\beta^*$. The shaded area represents the standard deviation. \label{fig:separation}}
\end{figure}

We now discuss the part (2) of the definition of spatial hierarchy, that is to say how the hubs are located in space. Indeed, another property that we can expect from spatially hierarchical graph is that of \emph{geographical separation}, as defined in the section Material and Methods. We define a separation index (see the section Methods for the definition) which quantifies the separation between the respective influence zones of hubs belonging to the same level. The separation function is equal to $1$ if the nodes' influence zones do not overlap at all and 0 if they perfectly overlap. We plot this quantity averaged over the all the graph's levels for different values of $\beta/\beta^*$ on Fig.~\ref{fig:separation}. One can observe on this graph that the separation index reaches values above $0.90$ when $\beta/\beta^* \geq 1$, which means that the corresponding graphs indeed have a structure with hubs controlling geographically well-separated regions. Obviously, the choice of the shape of the influence zone (which is chosen here to be a disk, see Materials and Methods) strongly impacts the results but the same qualitative behavior will be obtained for any type of convex shapes.

In conclusion, the graphs produced by our model in the regime $\beta / \beta^*$ satisfy the two points of the definition. They exhibit a spatially hierarchical structure, characterised by a distance ordering and geographical separation of hubs. We saw earlier that in this regime we have specific, non trivial properties such as $L_{tot}$ scaling with an exponent depending continuously on $\beta/\beta^*$. Using a simple toy model, we will now show that the spatial hierarchy can explain this property.

\subsubsection{Understanding the scaling with a toy model} 

\begin{figure}[!h]
\centering
\includegraphics[width=0.30\textwidth]{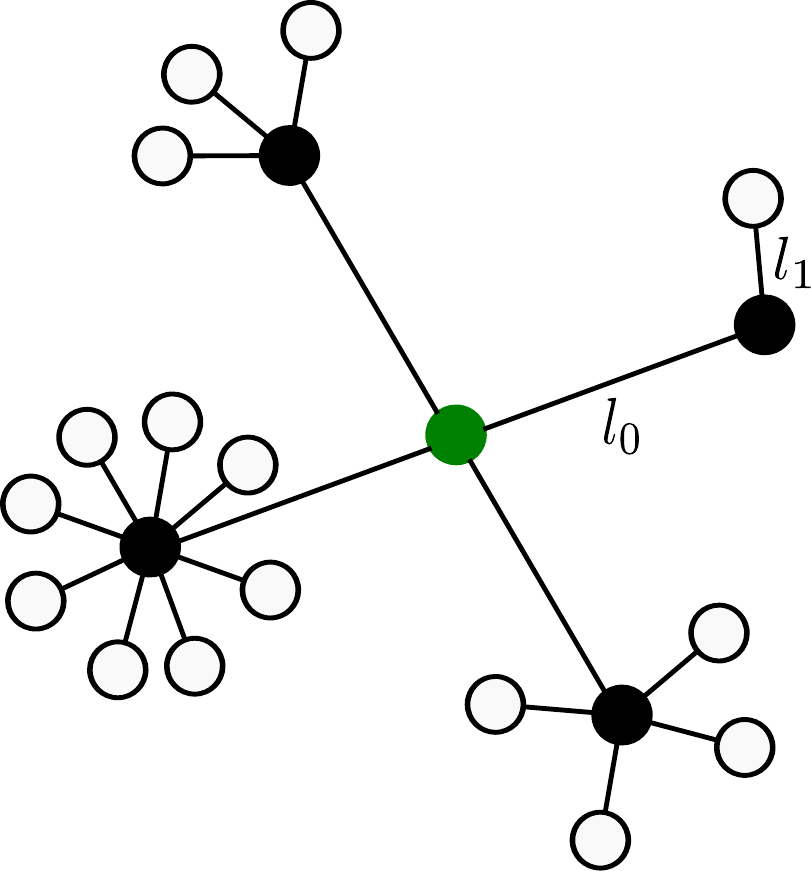}
\caption{A schematic representation of the hierarchical fractal network used as a toy model. 
\label{fig:fractal_network}}
\end{figure}

We consider the toy model defined by the fractal tree depicted on Fig.~\ref{fig:fractal_network} for which the distance between the levels $n$ and $n+1$ is given by
\begin{equation}
\ell_n=\ell_0 b^n
\label{eq:fractal_distance}
\end{equation}
where $b\in [0,1]$ is the scaling factor. Each node at the level $n$ is connected to $z$ nodes
at the level $n+1$ which implies that
\begin{equation}
N_n = z^n
\label{eq:fractal_nodes}
\end{equation}
where $z>0$ is an integer. A simple calculation on this graph shows that in the limit $z^g \gg 1$, the total length of the graph with g levels scales as
\begin{equation}
L_{tot} \sim N^{\frac{\ln(b)}{\ln(z)}+1}
\end{equation}
where $\frac{\ln(b)}{\ln(z)} +1 \leq 1$ because $b \leq 1$ and $z>1$. This simple model thus provides a simple mechanism accounting for continuous values of $\tau$ whose value depends on the scaling factor $b$. It provides a simplified picture of the graphs in the intermediate regime $\beta \simeq \beta^*$ and exhibits the key features of the graphs in this regime: the hub structure reminiscent of the star graph and where the nodes connected to each hub form geographically distinct regions, organized in a hierarchical fashion. It is also interesting to note that the parameter $z$ can be easily determined from the average degree of the network, and that the parameter $b$ of the toy model can be related to our model by measuring the decrease of the mean distance between different levels of the hierarchy, as in Fig.~\ref{fig:distance_hierarchy}. By plotting these curves for different values of $\beta/\beta^*$, we find that the coefficient of the exponential decays decreases linearly with $\beta / \beta^*$ and therefore that $b \sim e^{\beta/\beta*}$ (However, the comparison only makes sense in the regime $\beta \sim \beta^*$, as otherwise the graphs do not exhibit spatial hierarchy).

\subsection{Efficiency}

 Most transportation networks are not obtained by a global optimization but result from the addition of various, successive layers. The question of the efficiency of these self-organized systems is therefore not trivial and deserves some investigation. The model considered here allows us to test the effect of various parameters and how efficient a self-organized system can be. In particular, we would like to characterize the efficiency of the system for various values of $\beta$. For this, we can assume that the construction cost per unit length is fixed (ie. the factor $\eta$ in Eq.~\ref{eq:cost} is constant), and since $\beta = \frac{\eta}{\kappa}$ a change of value for $\beta$ is equivalent to a change in the benefits per passenger per unit of length. 

A first natural measure of how optimal the network is, is given by its total cost proportional to the total length $L_{tot}$: the shorter a network is, the better for the company in terms of building and maintenance costs. In our model, the behaviour of the total cost is simple and expected: for small values of $\beta/\beta^*$, the obtained networks correspond to a situation where the users are charged a lot compared to the maintenance cost, and the network is very long ($L_{tot}\propto N$). In the opposite case, when $\beta/\beta^* \gg 1$ the main concern in building this network is concentrated on construction cost and the network has the smallest total length possible (for a given set of nodes). 

\begin{figure}[!h]
\centering
\includegraphics[width=0.45\textwidth]{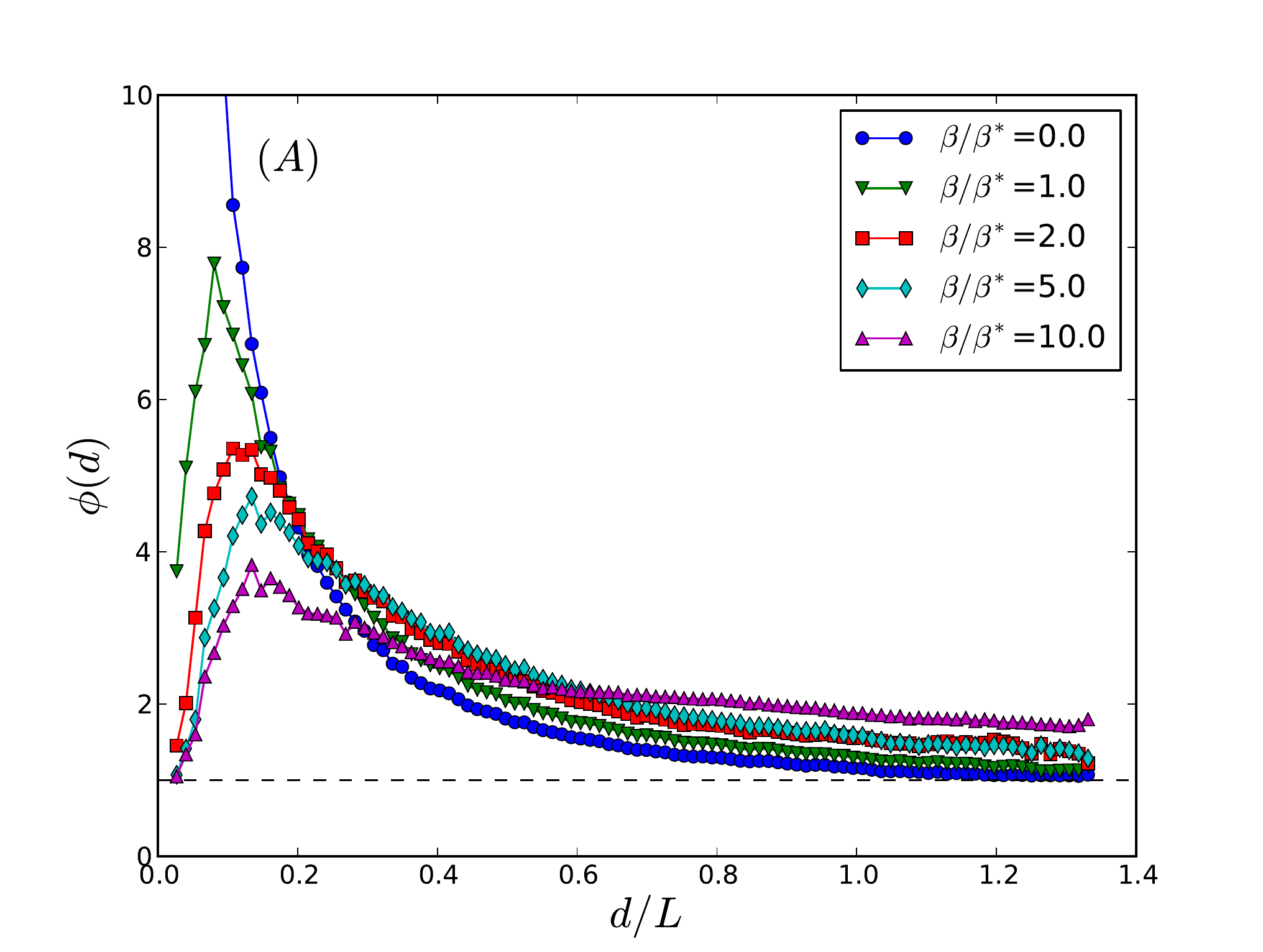}\\
\includegraphics[width=0.45\textwidth]{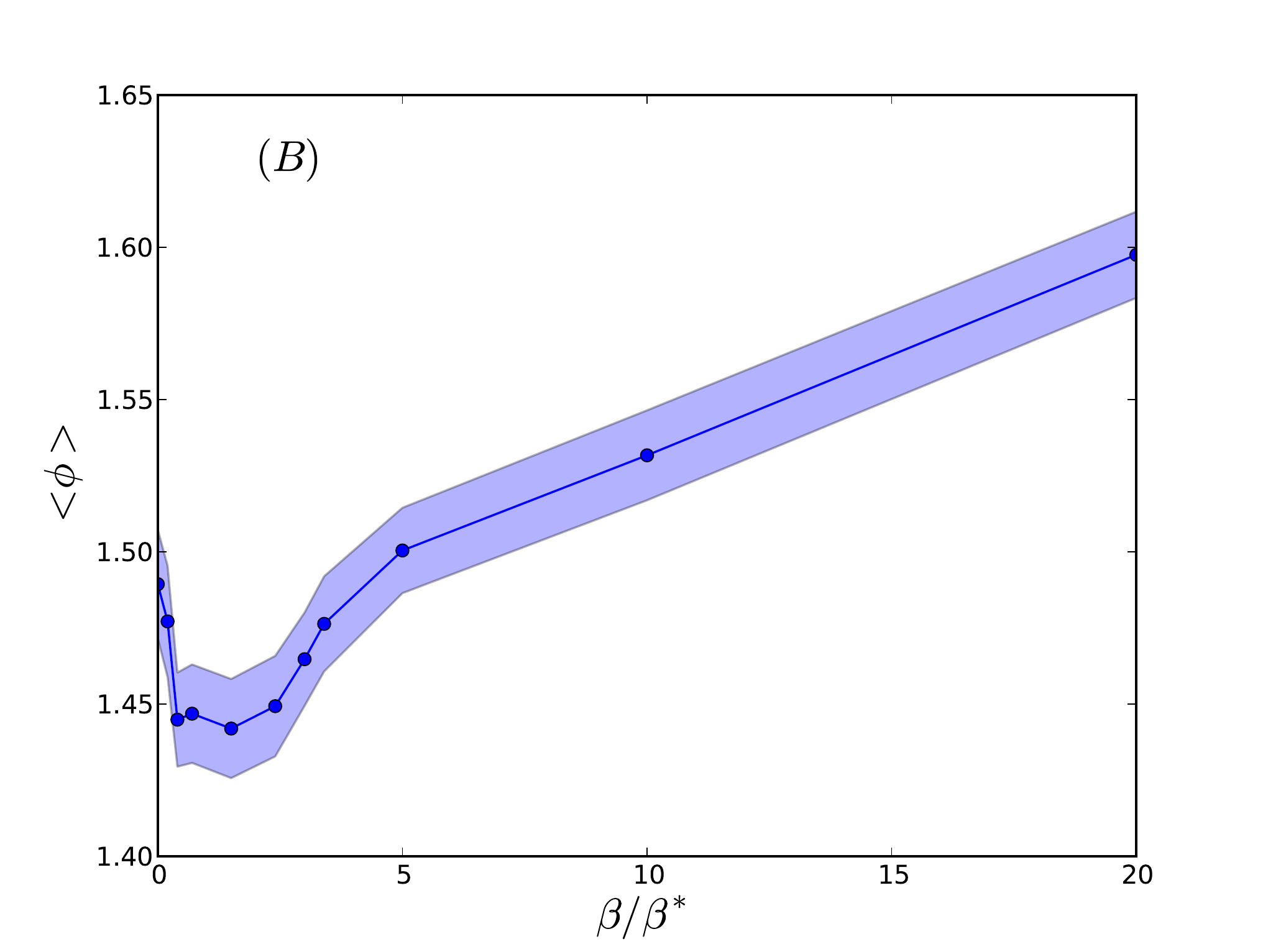}
\caption{(A) Detour function $\phi (d)$ versus the relative distance between nodes for different values of $\beta/\beta^*$. 
(B) Average detour index $< \phi >$  for several realisations of the graphs as a function of $\beta/\beta^*$. The shaded area represents the standard deviation of $< \phi >$. This plot shows that there is a minimum for this quantity in the intermediate regime $\beta\sim\beta^*$.
\label{fig:RLE}}
\end{figure}

The cost is however not enough to determine how efficient the network is from the users' point of view: a very low-cost network might indeed be very inefficient. A simple measure of efficiency is then given by the amount of detour needed to go from one point to another. In other words, a network is efficient if the shortest path on the network for most pairs of nodes is very close to a straight line. The detour index for a pair of nodes $(i,j)$ is conveniently measured by $D(i,j)/d(i,j)$ where $D(i,j)$ is the length of the shortest path between $i$ and $j$, and $d(i,j)$ is the euclidean distance between $i$ and $j$. In order to have a detailed information about the network, we use the
quantity introduced in~\cite{Aldous:2010}
\begin{equation}
\phi(d) = \frac{1}{\mathcal{N}(d)} \sum_{\substack{i,j\\d(i,j) =d}} \frac{D(i,j)}{d(i,j)}
\end{equation}
where the normalisation $\mathcal{N}(d)$ is the number of pairs with $d(i,j)=d$.  We plot this `detour function' for several values of $\beta/\beta^*$ on Fig.~\ref{fig:RLE}(A). For $\beta/\beta^* \ll 1$, the function $\phi(d)$ takes high values for $d$ small and low values for large $d$, meaning that the corresponding networks are very inefficient for relatively close nodes while being very efficient for distant nodes. On the other hand, for $\beta/\beta^* \gg 1$ we see that the MST is very efficient for neighboring nodes but less efficient than the star-graph for long distances. Surprisingly, the graphs for $\beta/\beta^* \sim 1$ exhibit a non trivial behaviour: for small distances, the detour is not as good as for the MST, but not as bad as for the star graph and for long distances it is the opposite. In order to make this statement more precise we compute the average of $\phi(d)$ over $d$ (a quantity which has a clear meaning for trees, see~\cite{Aldous:2010} for objections to the use of $< \phi(d) >$ as a good efficiency measure in general), and plot it as a function of $\beta/\beta^*$. The results are shown in Fig.~\ref{fig:RLE}(B) and confirm this surprising behavior in the intermediate regime: we observe a minimum for $\beta/\beta^* \sim 1$. In other words, there exists a non trivial value of $\beta$, i.e. a value of the benefits per passenger per unit of length, for which the network is optimal from the point of view of the users. 

\begin{figure}[!h]
\centering
\includegraphics[width=0.45\textwidth]{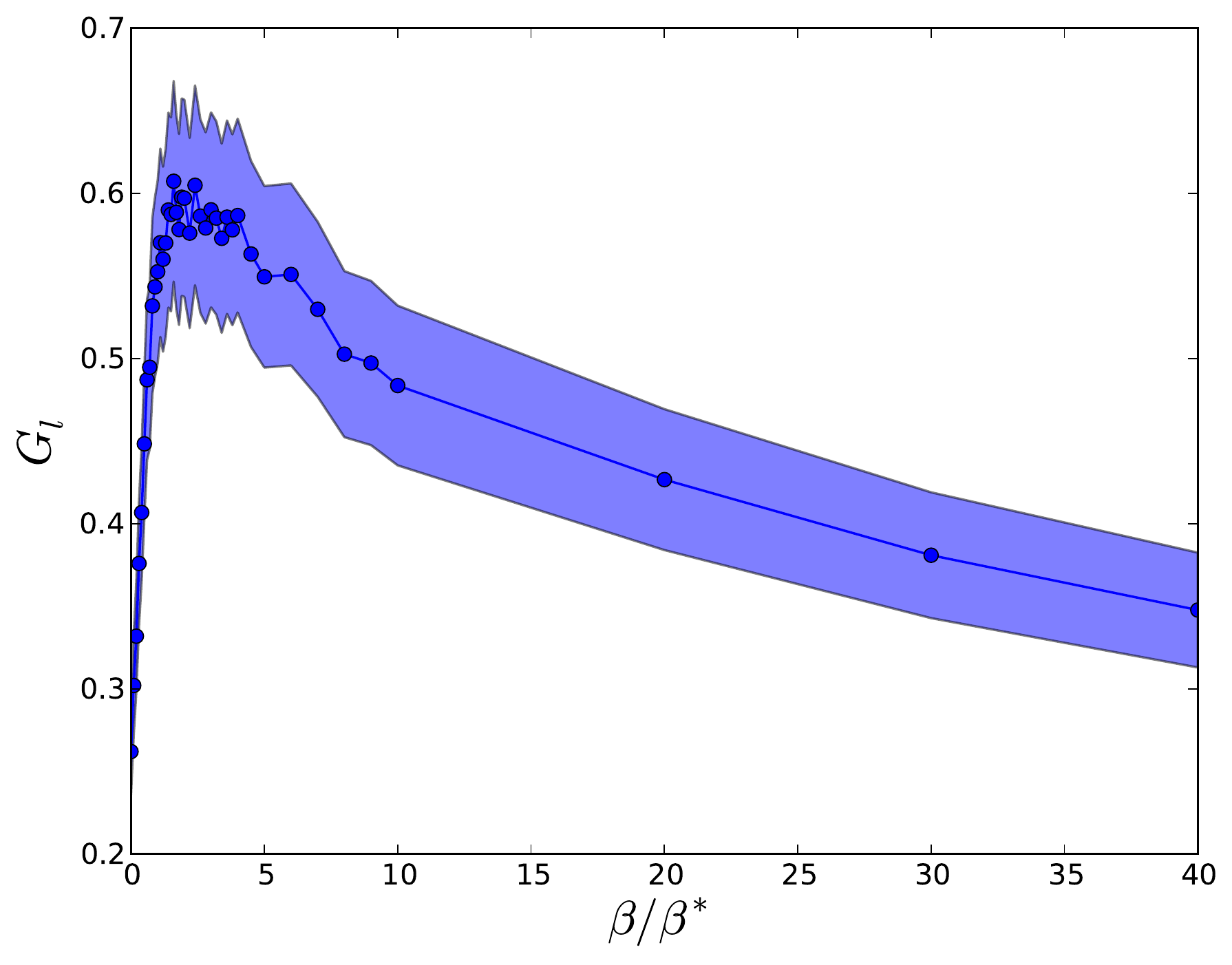}
\caption{Evolution of the Gini coefficient for the length versus $\beta/\beta^*$ (for different values of $\beta^*$). The shaded area represents the standard deviation. \label{fig:gini_length}}
\end{figure}

The existence of such an optimum is far from obvious and in order to gain more understanding about this phenomenon, we plot the Gini coefficient $G_l$ relative to the length of the edges between nodes in Fig.~\ref{fig:gini_length}. We observe that the Gini coefficient peaks around $\beta/\beta^* = 1$, which means that in this regime, the diversity in terms of edge length is the highest. The large diversity of lengths explains why the network is the most efficient in this regime: indeed long links are needed to cover large distances, while smaller links are needed to reach efficiently all the nodes. It is interesting to note that this argument is similar to the one proposed by Kleinberg \cite{Kleinberg:2000} in order to explain the existence of an optimal delivery time in small-world networks.

\section{Discussion}

We have presented a model of a growing spatial network based on a cost-benefit analysis. This model allows us to discuss the effect of a local optimization on the large-scale properties of these networks. First, we showed that the graphs exhibit a crossover between the star-graph and the minimum spanning tree when the relative importance of the cost increases. This crossover is characterized by a continuously varying exponent which could give some hints about other quantities observed in cities such as the total length travelled by the population. Secondly, we showed that the model predicts the emergence of a spatial hierarchical structure in the intermediate regime where costs and benefits are of the same order of magnitude. We showed that this spatial hierarchy can explain the non trivial behaviour of the total length versus the number of nodes. Finally, this model shows that in the intermediate regime the vast diversity of links lengths entails a large efficiency, an aspect which could of primary importance for practical applications.

An interesting playground for this model is given by railways and we can estimate the value of $\beta/\beta^*$ for these systems. In some cases, we were able to extract the data from various sources (in particular financial reports of railway companies) and the results are shown in Table~1. We estimate for different real-world networks, including some of the oldest railway systems, $\beta$ using its definition (total maintenance costs per year divided by the total length and by the average ticket price per km). In order to estimate $\beta^*$ we use Eq.\ref{eq:beta*_traffic} in the following way
\begin{equation}
\beta^*\simeq\frac{T_{tot}}{L_{tot}}
\end{equation}
where $T_{tot}$ is the total travelled length (in passengers$\cdot$kms$/$year) and $L_{tot}$ is the total length of the network under consideration. Remarquably, the computed values for the ratio $\beta/\beta^*$ shown in Table~1 are all of the order of $1$ (ranging from $0.20$ to $1.56$). In the framework of this model, this result shows that all these systems are in the regime where the networks possess the property of spatial hierarchy, suggesting it is a crucial feature for real-world networks. We note that in our model, the value of $\beta / \beta^*$ is given exogeneously, and it would be extremely interesting to understand how we could construct a model leading to this value in an endogeneous way.

\begin{table*}
\caption{{\bf Empirical estimates for $\beta$ and $\beta^*$}. Table giving the total ride distance (in km), the total network length (in km), the total annual maintenance expenditure (in euros per year)  and the average ticket price (in euros per km). All the given values correspond to the year 2011. From these data we compute the experimental values of $\beta$, $\beta^*$ and their ratio (data obtained from various sources such as financial reports of railway companies) \label{table:b_b*}}
\scalebox{0.8}{
\begin{tabular}{lccccccc}
\hline
&&&&&&&\\
Country & Total distance travelled & Total length & $\beta^*$ & Maintenance cost  & Average ticket price & $\beta$ & $\beta/\beta^*$ \\
 & \footnotesize(passenger$\cdot$kms$/$year) & \footnotesize(kms) & \footnotesize(passengers$/$year) & \footnotesize(euros$/$year) & \footnotesize(euros$/$km) & \footnotesize(passengers$/$year) & \\
\hline
France & $88.1 \, 10^9$ & $29,901$	& $2.94 \, 10^6$ &	$2.10 \, 10^9 $ & $0.12$ & $5.85 \, 10^5$ & $\mathbf{0.20}$ \\
Germany & $79.2 \, 10^9$ & $37,679$	& $2.10 \, 10^6$ &	$7.50 \, 10^9 $ & $0.30$ & $6.60 \, 10^5$ & $\mathbf{0.32}$ \\
India & $978.5 \, 10^9$ & $65,000$	& $1.51 \, 10^7$ &	$3.00 \, 10^9 $ & $0.01$ & $4.61 \, 10^6$ & $\mathbf{0.31}$ \\
Italy & $40.6 \, 10^9$ & $24,179$	& $1.68 \, 10^6$ &	$4.30 \, 10^9 $ & $0.20$ & $8.89 \, 10^5$ & $\mathbf{0.53}$ \\
Spain & $22.7 \, 10^9$ & $15,064$	& $1.51 \, 10^6$ &	$3.16 \, 10^9 $ & $0.11$ & $1.91 \, 10^6$ & $\mathbf{1.26}$ \\
Switzerland & $18.0 \, 10^9$ & $5,063$	& $3.55 \, 10^6$ &	$2.03 \, 10^9 $ & $0.17$ & $2.36 \, 10^6$ & $\mathbf{0.66}$ \\
United Kingdom & $62.7 \, 10^9$ & $16,321$	& $3.84 \, 10^6$ &	$12 \, 10^9 $ & $0.16$ & $4.59 \, 10^6$ & $\mathbf{1.19}$ \\
United States & $17.2 \, 10^9$ & $226,427$	& $7.59 \, 10^4$ &	$2.96 \, 10^9 $ & $0.11$ & $1.18 \, 10^5$ & $\mathbf{1.56}$\\
\hline
\end{tabular}
}
\end{table*}

There are also several directions that seem interesting. First, various forms of cost and benefits functions could be investigated in order to model specific networks. In particular, there are several choices that can be taken for the expected traffic. In this paper we limited ourselves to estimate the traffic as a direct traffic from a node $i$ to a node $j$, but it is likely that part of the traffic will come from other nodes. In order to take this into account, we think that the following extensions are probably interesting:
\begin{enumerate}
\item A given city (denoted by $0$ with population $M_0$) plays a particular role in the network (the capital city in a relatively small country, for example). In that case it is beneficial to be close to that city through the network and we write
\begin{equation}
\label{eq:R1}
R^{(1)}_{ij} = (1-\lambda)\frac{M_i M_j}{d_{ij}^{a-1}}  + \lambda \: \frac{M_i M_0}{\left(D_{0j} + d_{ij}\right)^{a-1}}- \beta \: d_{ij}
\end{equation}
where $\lambda \in \left[ 0,1 \right]$ is a coefficient weighing the relative importance of the traffic coming from the particular city.
\item The most general case where all the network-induced traffic are taken into account. We then consider
\begin{equation}
\label{eq:R2}
R^{(2)}_{ij} = \sum_{k \neq i}\frac{M_i M_k}{\left(D_{kj} + d_{ij}\right)^{a-1}}- \beta \: d_{ij}
\end{equation}
\end{enumerate} 
Other ingredients such as the presence of different rail companies, or the difference between a state-planned network and a network built by private actors, etc, could easily be implemented and the corresponding models could possibly lead to interesting results.

More importantly, we limited ourselves here to trees in order to focus on the large-scale consequences of the cost-benefit mechanism. Further studies are needed in order to uncover the mechanisms of formation of loops in growing spatial networks and we believe that the model presented here might represent a suitable modeling framework.

Finally, it seems plausible that the general cost-benefit framework introduced at the beginning of the article could be applied to the modelling of systems besides transportation networks. We believe it captures the fundamental features of spatial network while being versatile enough to model the growth of a great diversity of systems shaped by space.

\section{Materials and Methods}

\subsection{Simulations} 

The simulation starts by distributing nodes uniformly in a square. We then attribute to each node a random population distributed according to the power law
\begin{equation}
P_M(x) = \frac{\mu}{x^{\mu+1}}
\end{equation}
The choice of this distribution is motivated by Zipf's empirical results on city populations~\cite{Zipf:1949} (which motivates the choice $\mu=1.1$ in our simulations) but also because we can go from a peaked to a broad distribution by tuning the value of $\mu$. Indeed, for $\mu>2$, both the first and the second moment of this distribution exist and the distribution can be considered as peaked. In contrast for $1<\mu<2$, only the first moment converges and the distribution is broad.

Once the set of nodes is generated, we choose a random node as the root and add nodes recursively until all the nodes belong to the graph. At each time step, the nodes belonging to the graph constitute the set of `inactive nodes', and the other -not yet connected - nodes the `active' nodes. At each time step we connect an active node to an inactive node such that their value of $R$ defined in Eq.~\ref{eq:R0} is maximum.

\subsection{Geographic separation} 

\begin{figure}[!h]
\centering
\includegraphics[width=0.45\textwidth]{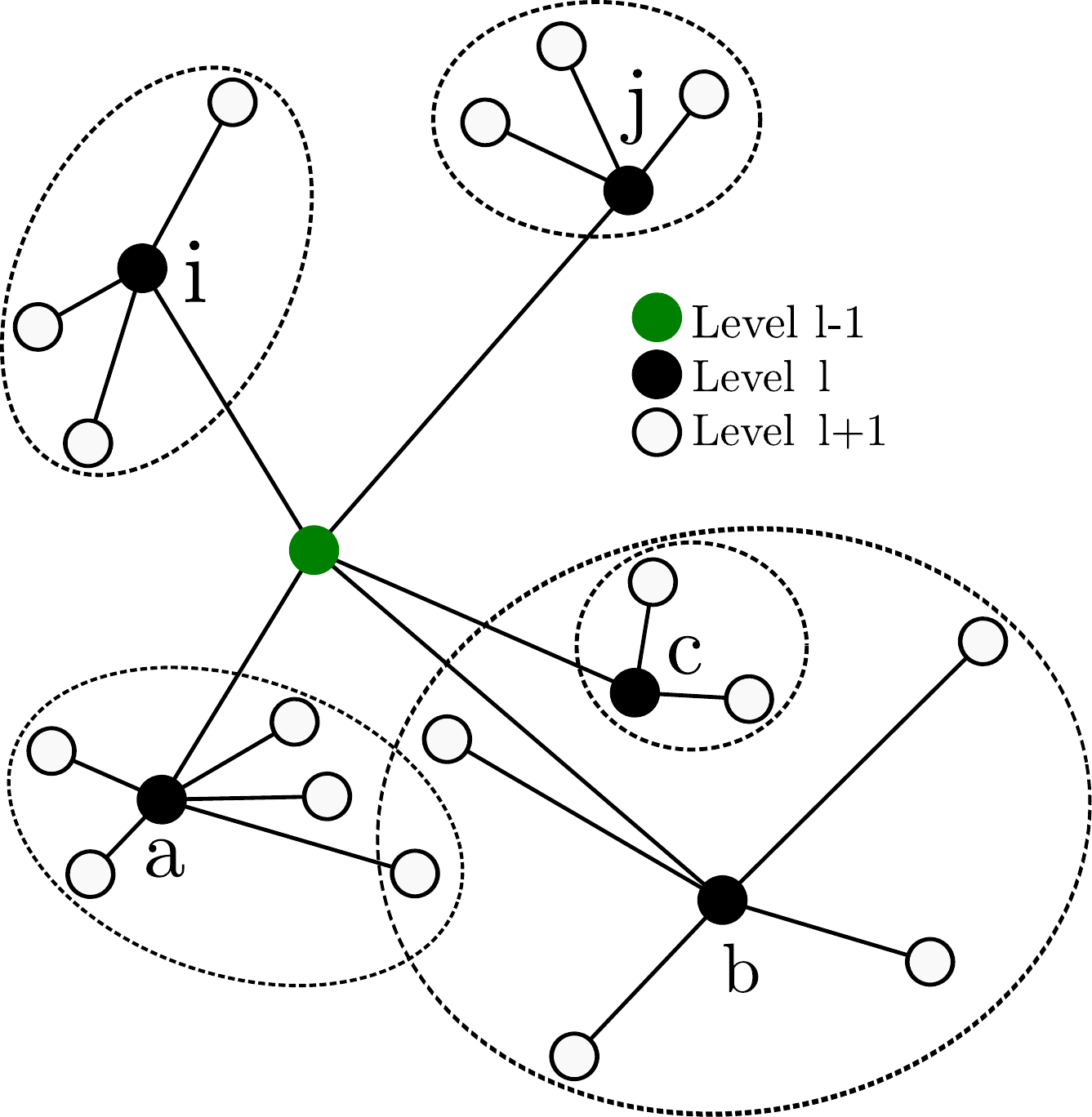}\\
\caption{Illustration of the influence zones (dotted lines) around several hubs. We have, according to the definition of the separation index, $S(i,j)=0$, $0 < S(a,b)< 1$ and $S(b,c)=1$.
\label{fig:separation_illustration}}
\end{figure}

We say that a graph is geographically separated if the influence zones of every node of a given hierarchical level do not overlap and if they are included in the influence zone of the nodes of the previous level in the hierarchy. Formally, if we designate by $\mathcal{I}^i_l$ the influence zone of the node $i$ located at level $l$ in the hierarchy, $\mathcal{I}_l = \cup_{i \in l} \mathcal{I}^i_l$ the reunion of all the influence zones for nodes belonging to the level $n$. We say that the graph is geographically separated if:
\begin{align}
&\mathcal{I}_l \subset \mathcal{I}_{l+1} \; \forall l\\
&\mathcal{I}^i_l \cap \mathcal{I}^j_l = \O \; \text{if} \; j \neq i, \forall l
\end{align}
The degree of geographical separability of a graph strongly depends on the definition of the influence zone of a node. For instance, if we take the influence zone of a node $i$ to be the surface of smallest area containing all the nodes connected to $i$, it follows that all planar graph are totally separated. 
In the context of transportation networks, we expect hubs to radiate up to a certain distance around them, that is to say connect to all the nodes located in a convex shape. We simply define the influence zone of a node $i$ as the circle centered on the barycenter of i's neighbours that belong to the next level, of radius the maximum distance between the barycenter and those points. 
Figure~\ref{fig:separation_example} is intended to help the reader visualise these influence zones on an example: The green circle represent the influence zone of the root and the red circles the influence zones of the hubs connected to it. One can see that the graph is geographically separated up to a good approximation.

In order to quantify this notion of geographical separability, we define the separation index of the level $l$ as the average over all the nodes belonging to $l$ of the separation function. The separation function is equal to $1$ if the distance $d(i,j)$ between the centers of the influence zones of $i$ and $j$ is larger than their respective radius (no overlap), and equal to 
\begin{equation}
S(i,j) = 1-\frac{\text{Area of the overlap between} \; \mathcal{I}_l^i \; \text{and} \; \mathcal{I}_l^j}{\min \left(\text{Area of} \; \mathcal{I}_l^i\text{, Area of} \; \mathcal{I}_l^j\right)}
\end{equation}
One can see that the separation function is equal to 1 if the nodes' influence zones do not overlap at all and 0 if they perfectly overlap (all the influence zones overlapping, like Russian dolls). Therefore, the separation index is equal to 1 if the level s is perfectly separated and 0 if the influence zones are completely mixed. One can see on Fig.~\ref{fig:separation_illustration} an illustration expliciting the value of the separation index for different situations.

\section{Acknowledgments}
PJ thanks A. Bonnafous, G. Michaud, and C. Raux for discussions at an early stage of the project.

\section{Notes}

RL, PJ, MB designed, performed research and wrote the paper.

\bibliographystyle{prsty}

\end{document}